\documentclass[conference]{IEEEtran}
\IEEEoverridecommandlockouts
\usepackage{flushend}
\usepackage{caption}
\usepackage{subcaption}
\usepackage{graphicx}
\usepackage[table, svgnames, dvipsnames]{xcolor}
\usepackage{booktabs,makecell}
\usepackage{comment}
\usepackage{cite}
\usepackage{amsmath,amssymb,amsfonts}
\usepackage{algorithmicx}
\usepackage{algcompatible}
\definecolor{green}{rgb}{0.16, 0.67, 0.53}
\usepackage{textcomp,balance,url}
\usepackage[left=1.62cm,right=1.62cm,top=1.9cm]{geometry}
\definecolor{chestnut}{rgb}{0.97, 0.51, 0.47}
\def\BibTeX{{\rm B\kern-.05em{\sc i\kern-.025em b}\kern-.08em
    T\kern-.1667em\lower.7ex\hbox{E}\kern-.125emX}}

\usepackage{capt-of}
\usepackage{algorithm}
\usepackage{algpseudocode}
\usepackage{dblfloatfix}
\usepackage{fixltx2e}

\usepackage{capt-of}

\definecolor{blue}{rgb}{0.0, 0.47, 0.75}
\begin{document}

\title{Predictive RTO for CoAP using Lightweight Support Vector Regression in Internet of Things}

\author{
\IEEEauthorblockN{Tobias Hansson and Praveen Kumar Donta}

\IEEEauthorblockA{\IEEEauthorrefmark{1}\textit{Department of Computer Systems and Sciences}, \textit{Stockholm University,} Stockholm 164 25, Sweden.} 
% \IEEEauthorblockA{\IEEEauthorrefmark{3}\textit{Distributed Systems Group, TU Wien}, Vienna 1040, Austria and \textit{ICREA} Barcelona 08002, Spain}
 \texttt{tobiashansson.01@gmail.com, praveen@dsv.su.se}}

\maketitle

\begin{abstract}
Internet of Things (IoT) networks require lightweight application layer messaging, and CoAP is an option because it supports REST-style interactions over UDP on constrained devices. However, CoAP congestion control still depends on fixed heuristics, including binary exponential backoff (BEB) and RTT-based mechanisms such as CoCoA and CoCoA+, which do not adapt well to dynamic and lossy wireless links. This paper proposes \texttt{prCoAP}, a lightweight data-driven approach that replaces heuristic Retransmission Timeout (RTO) selection with a per-attempt linear Support Vector Regression (SVR) ensemble for direct RTO prediction from node-observable features. The model runs on-device on low-end microcontrollers and operates within strict memory and energy budgets. The framework also includes a calibrated Random Forest drop classifier that identifies likely-to-fail transactions in later retransmission attempts and terminates them early to reduce channel occupancy. We evaluate the approach using a discrete-event simulator implementing IEEE 802.15.4 and RFC 7252 and validate it against the FIT IoT-LAB testbed. Our experiments confirm that the proposed linear SVR achieves 97.25\% PDR, outperforming standard CoAP under the evaluated conditions. We also evaluate a kernel SVR variant; while it improves regression fit (R2 0.84 vs. 0.63), the linear SVR provides better system-level efficiency, achieving comparable PDR with lower energy overhead.
\end{abstract}


\section{Introduction}\label{sec:Introduction}
\IEEEPARstart{I}{nternet of Things}  (IoT) connects physical devices with embedded sensing, actuation, and communication so they can exchange data over networks \cite{li2015internet}. IoT applications such as smart grids, industrial automation, agriculture, and healthcare  often rely on low-cost embedded nodes with limited memory, limited energy budget, and low-power wireless links such as IEEE 802.15.4 that can be lossy and unstable \cite{Alaa2025towards}. In these constrained settings, conventional Internet protocols such as HTTP are inefficient because they requires per-connection state, window management, and a three-way handshake, which increase processing and memory overhead and reduce suitability for constrained nodes \cite{donta2023towards,aveleira2025coap_uad,donta2022survey,lopez6721637coap}. 

The IETF standardized the Constrained Application Protocol (CoAP) in RFC 7252 as a lightweight application layer protocol that runs over UDP and supports a REST-style request/response model \cite{rfc7228} to address the above-mentioned gaps. CoAP provides reliability at the application layer using Confirmable messages, Acknowledgments, and a Retransmission Timeout (RTO) that controls when an unacknowledged message is retransmitted, making RTO estimation a key factor for congestion behavior and network efficiency. The default CoAP specification selects an initial RTO randomly between 2-3 s and applies Binary Exponential Back-off (BEB) by doubling the RTO for up to four retransmissions without adapting to observed network conditions. This static mechanism cannot distinguish congestion-induced loss from wireless link errors, which can reduce utilization when the channel is lossy but not congested and can cause retransmission bursts under heavy load \cite{donta2023icocoa}. 

\begin{table}[b]
\centering
\caption{A Summary of CoAP RTO calculation strategies from the literature.}
\label{tab:rto_comparison}
\resizebox{\columnwidth}{!}{%
\begin{tabular}{lccccc}
\toprule
\textbf{Method} &
\textbf{\makecell{Direct\\RTO\\Pred.}} &
\textbf{\makecell{Node-\\Side\\Only}} &
\textbf{Lightweight} &
\textbf{\makecell{Data-\\Driven}} &
\textbf{\makecell{PDR\\Gain}} \\
\midrule
CoCoA \cite{betzler2016coap}          & No  & Yes & Yes & No  & Moderate \\
CoCoA+ \cite{betzler2015cocoa}    & No  & Yes & Yes & No  & Moderate \\
FASOR \cite{jarvinen2018fasor}          & No  & Yes & Yes & No  & Moderate \\
CACC \cite{akpakwu2020cacc}            & No  & Yes & Yes & No  & Moderate \\
ML-CoCoA \cite{demir2020mlcocoa}   & No  & Yes & Yes & Yes & Low \\
iCoCoA \cite{donta2023icocoa}        & Yes & No  & No  & Yes & High \\
Proposed (SVR)           & Yes & Yes & Yes & Yes & High \\
\bottomrule
\end{tabular}%
}
\end{table}
Recent research has improved CoAP congestion control beyond the default static RTO and BEB  as summarized in Table~\ref{tab:rto_comparison}. For example, CoCoA uses dual SRTT estimators, a strong estimator from unambiguous ACKs and a weak estimator from retransmission-related ACKs, and applies a Variable Backoff Factor (VBF) to limit unnecessary RTO growth \cite{betzler2016coap}. CoCoA+ adds RTO ageing, stronger SRTT filtering ($\alpha = 0.05$), and a higher weak-estimator weight to improve behavior under sustained impairment \cite{betzler2015cocoa}. FASOR separates RTT samples into fast and slow paths to reduce buffer-bloat effects in high congestion \cite{jarvinen2018fasor}, while CACC incorporates retransmission context and labels samples as strong, weak, or failed to better differentiate loss causes \cite{akpakwu2020cacc}.  Despite these advances, heuristic designs often do not generalize well across dynamic IoT conditions, remain reactive to past RTT, and typically do not explicitly leverage multivariate local state for proactive RTO selection. ML approaches exist \cite{demir2020mlcocoa, jiang2021machine, donta2023icocoa}, but many only tune heuristic parameters \cite{demir2020mlcocoa}, while others require computationally impractical efficient protocols for constrained nodes \cite{donta2023icocoa}, leaving a gap for a lightweight, on-device ML RTO predictor. %A summary on CoAP RTO approaches shown in Table~\ref{tab:rto_comparison}. 

To address the above limitations, we propose a lightweight predictive RTO for CoAP (\texttt{prCoAP}) using Support Vector Regression (SVR) that predicts an absolute RTO directly from node-observable state, without heuristic coefficient tuning or reliance on global network information. In this context, the main contributions are as follows:
\begin{itemize}
\item We design an on-device RTO predictor based on a per-attempt linear SVR ensemble that uses five observable features (SRTT, RTTVAR, attempt index, inter-arrival time, packet success rate) to predict a dynamic RTO.
\item We develop a learning pipeline using a log-domain RTO transformation and feature standardization, improving prediction accuracy under skewed timeout distributions while keeping inference to a single dot product and a small memory footprint suitable for resource-constrained devices.
\item We add a calibrated transaction-drop classifier activated from the third retransmission attempt to trigger early abandonment under severe degradation, reducing channel occupancy and energy waste.
\item We validate our \texttt{prCoAP} with a cost-aware evaluation that accounts for inference overhead and with real-hardware experiments on FIT IoT-LAB\footnote{\url{https://www.iot-lab.info/}}, benchmarking against baselines across varied client concurrency, delivery regimes, traffic patterns, and payload sizes.
\end{itemize}
The remaining sections of this paper are organized as follows. The proposed \texttt{prCoAP} protocol re-design is discussed in Section~\ref{sec3}. A detailed discussion of the experiments and numerical results are provided in Section~\ref{sec4}. We conclude our paper with future scope in Section~\ref{sec5}.

\section{Proposed Predictive RTO for CoAP}\label{sec3}
This section describes the proposed \texttt{prCoAP} design, the model training and validation process, and the Random Forest drop policy used for early transaction termination.
\subsection{Node-Observable States}
\begin{figure}[t]
    \centering
    \includegraphics[width=0.5\textwidth]{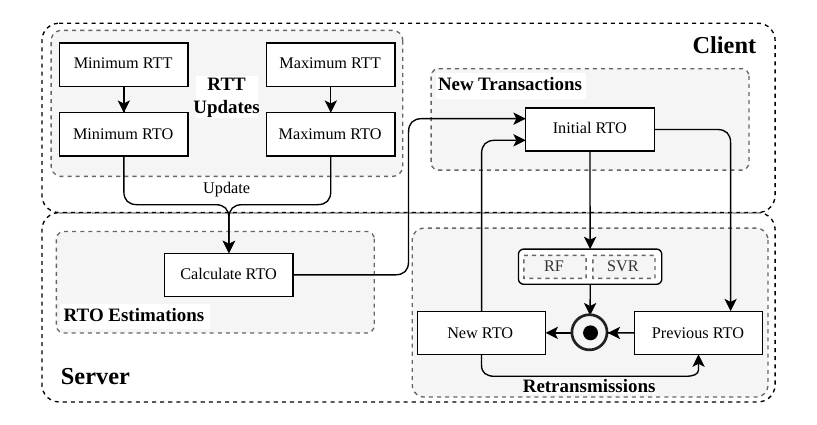}
    \caption{General architecture of the proposed Predictive RTO for CoAP}
    \label{fig:generalarch}
\end{figure}

The proposed \texttt{prCoAP} replaces traditional CoAP's heuristic RTO selection with a direct, node-side prediction of an absolute retransmission timeout (RTO) computed from locally observable state as shown in Fig. \ref{fig:generalarch}. The proposed method operates inside the standard CoAP CON/ACK loop. For each transmission attempt, the node (i) updates its RTT estimators from prior ACKs, (ii) constructs a compact feature vector, (iii) selects an attempt-specific Linear SVR model to predict the next RTO, and (iv) from later attempts, optionally invokes a calibrated drop classifier to abandon transactions predicted to timeout, reducing channel occupancy and energy waste. The approach uses only information available at the sender and keeps inference lightweight by using linear models with a small parameter footprint. Specifically, \texttt{prCoAP} computes the RTO from a compact set of timing and reliability signals already available in the CoAP sender state. At time $t$, the node forms the feature vector as shown in Eq.~(\ref{eq1}).
\begin{equation}\label{eq1}
\mathbf{x}_t = \big[ \mathrm{SRTT}_t,\ \mathrm{RTTVAR}_t,\ a_t,\ \mathrm{IAT}_t,\ \mathrm{SR}_t \big]^\top \in \mathbb{R}^5,
\end{equation}
where $\mathrm{SRTT}$ is the smoothed RTT, $\mathrm{RTTVAR}$ is RTT variance, $a_t$ is the attempt index, $\mathrm{IAT}_t$ is the inter-arrival time for the client traffic, and $\mathrm{SR}_t$ is a short-range success-rate estimate. The RTT estimators follow exponential updates:
\begin{equation}\label{eq:srtt}
\mathrm{SRTT} \leftarrow (1-\alpha)\mathrm{SRTT} + \alpha \cdot \mathrm{RTT}_{\text{sample}},
\end{equation}
\begin{equation}\label{eq:rttvar}
\mathrm{RTTVAR} \leftarrow (1-\beta)\mathrm{RTTVAR} + \beta\cdot \left| \mathrm{SRTT} - \mathrm{RTT}_{\text{sample}} \right|
\end{equation}
where the $\alpha$ and $\beta$ values assumed as 0.125, and 0.25, respectively \cite{betzler2016coap}. The $\mathrm{SR}_t$ is computed over a window of $W=10$ recent transactions as shown in Eq.(\ref{eq:sr})
\begin{equation}\label{eq:sr}
\mathrm{SR}_t = \frac{1}{W}\sum_{i=1}^{W}\mathbb{1}\{\text{success}_{t-i}\}.
\end{equation}
To reduce stale optimism after idle periods, the implementation decays the stored window score by a factor of $0.90$ per 2\,s idle interval.
\subsection{Direct RTO Prediction and Early Drop Control}
A single linear regressor trained across all attempts must cover a wide, right-skewed target range (approximately $100$\,ms to $60{,}000$\,ms), which degrades linear accuracy. The proposed \texttt{prCoAP} therefore trains an attempt-stratified Linear SVR ensemble.  We train and validate all models offline before integration. The simulator generates the dataset from a 300-scenario parameter sweep, and we split the data 80/20 using stratified sampling. We stratify the split by attempt index and channel regime to preserve the per-attempt target distribution. Each training sample corresponds to one transmission attempt, and the regression label $y$ is the attempt-level target RTO in milliseconds assigned for that attempt in the simulator after applying the attempt cap $S_a$. This procedure ensures that both training and evaluation samples are drawn from the same simulator distribution, while preserving the class balance required for the timeout label. The RF drop classifier achieves a ROC-AUC of 0.7186 and, at the selected operating threshold, attains 0.957 precision on the timeout class, which limits unnecessary early drops. The classifier is enabled only from the third retransmission attempt onward and applies a success-rate-conditioned decision threshold, so early termination occurs primarily under degraded channel conditions. For attempts $a\in\{0,\dots,6\}$, the target space is bounded by BEB-aligned caps $S_0=2000,\ S_1=4000,\ S_2=8000,\ S_3=16000,\ S_4=32000,\ S_5=S_6=60000\ \text{ms},$
and the node uses model $f_a(\cdot)$ at attempt $a$. In RFC~7252, the default setting is \texttt{MAX\_RETRANSMIT}=4; in our experiments we set \(A_{\max}=6\)  to examine late-stage retransmission behavior. Prior to training and inference, features are standardized using
\begin{equation}\label{eq:standardize}
\mathbf{z} = \frac{\mathbf{x}-\boldsymbol{\mu}}{\boldsymbol{\sigma}},
\end{equation}
and the target is log-transformed to reduce skew:
\begin{equation}
\tilde{y} = \log(1+y),\qquad
\hat{y} = \exp(\hat{\tilde{y}})-1.
\label{eq:logtarget}
\end{equation}
Each attempt-specific model predicts the log-domain target using a linear SVR
\begin{equation}
\hat{\tilde{y}} = f_a(\mathbf{z}) = \mathbf{w}_a^\top \mathbf{z} + b_a,
\label{eq:lin_svr}
\end{equation}
trained with the $\epsilon$-insensitive objective
\begin{align}
\min_{\mathbf{w},b,\boldsymbol{\xi},\boldsymbol{\xi}^*}\quad
&\frac{1}{2}\|\mathbf{w}\|^2 + C\sum_{i=1}^{n}(\xi_i+\xi_i^*) \label{eq:svr_obj}\\
\text{s.t.}\quad
& y_i - (\mathbf{w}^\top \mathbf{z}_i + b) \le \epsilon + \xi_i^*, \nonumber\\
& (\mathbf{w}^\top \mathbf{z}_i + b) - y_i \le \epsilon + \xi_i, \nonumber\\
& \xi_i,\xi_i^* \ge 0. \nonumber
\end{align}
Hyperparameters are selected per attempt using 3-fold cross-validated grid search $C \in \{0.01,0.1,1.0,10.0\}$, and $
\epsilon \in \{0.01,0.05,0.1,0.2\},$ optimizing $R^2$ in log space. Since inference is a single dot product over $d=5$ features, the runtime complexity is $O(d)$ per decision, and the stored parameter set remains small (During our experiments, it is approximately 768 bytes at double precision).

From the third retransmission attempt onward ($a\ge 3$), \texttt{prCoAP} additionally applies an early-abandonment policy using a calibrated Random Forest classifier. The classifier outputs a calibrated probability $p_{\text{to}}$ that the current transaction will ultimately timeout. The base model uses 100 trees, maximum depth 8, and \texttt{class\_weight="balanced"}, and the probabilities are calibrated using isotonic regression to support threshold-based decisions. \texttt{prCoAP} uses an adaptive threshold conditioned on the observed success rate:
\begin{equation}
\tau(\mathrm{SR})=
\begin{cases}
0.75, & \mathrm{SR}<0.30\\
0.92, & \mathrm{SR}>0.80\\
0.88, & \text{otherwise}
\end{cases}
\label{eq:tau}
\end{equation}
and drops when $p_{\text{to}} \ge \tau(\mathrm{SR})$. To reflect deployment uncertainty and avoid unrealistically brittle behavior, the implementation injects a residual false-drop probability of $0.01$ at inference time. Algorithm~\ref{alg:prcoap} summarizes the runtime integration of the SVR ensemble and the drop classifier inside the CoAP transaction loop.

\begin{algorithm}[t]
\caption{\texttt{prCoAP} runtime integration}
\label{alg:prcoap}
\begin{algorithmic}[1]
\Require SVR ensemble $\{f_a\}_{a=0}^6$, scaler $(\boldsymbol{\mu},\boldsymbol{\sigma})$, calibrated RF $g$
\Require max attempts $A_{\max}=6$, window length $W=10$
\For{each transaction}
\State $a\gets 0$
\While{$a\le A_{\max}$}
\State Form $\mathbf{x}\gets[\mathrm{SRTT},\mathrm{RTTVAR},a,\mathrm{IAT},\mathrm{SR}]^\top$
\State Standardize $\mathbf{z}\gets(\mathbf{x}-\boldsymbol{\mu})/\boldsymbol{\sigma}$
\State Predict $\hat{\tilde{y}}\gets f_a(\mathbf{z})$ and set $\widehat{\mathrm{RTO}}\gets \exp(\hat{\tilde{y}})-1$
\If{$a\ge 3$}
\State $p_{\text{to}}\gets g([\mathrm{SRTT},\mathrm{RTTVAR},\mathrm{SR},\mathrm{IAT}])$
\If{$p_{\text{to}} \ge \tau(\mathrm{SR})$}
\State Abort transaction; update failure window; \textbf{break}
\EndIf
\EndIf
\State Transmit CON; wait $\widehat{\mathrm{RTO}}$ for ACK
\If{ACK received}
\State Update $\mathrm{SRTT},\mathrm{RTTVAR}$ using \eqref{eq:srtt}--\eqref{eq:rttvar}; update success window; \textbf{break}
\Else
\State Update failure window; $a\gets a+1$
\EndIf
\State Update $\mathrm{SR}$ using \eqref{eq:sr}
\EndWhile
\EndFor
\end{algorithmic}
\end{algorithm}

\section{Performance Evaluation}\label{sec4}
This section describes the experimental setup and presents a detailed analysis of the numerical results.
\subsection{Setup}
All simulations and model training are implemented in Python. We use \textit{NumPy} and \textit{Pandas} for data processing, and \textit{scikit-learn} with \textit{joblib} for training and serializing the SVR and RF models. Evaluation parameters are summarized using Table \ref{tab:evaluation_params}. For reproducibility, we open source our simulations\footnote{\url{https://github.com/HHansssonn/ML_SVR_RTO_PREDICTION}}.
\begin{table}[t]
\centering
\caption{Evaluation parameters overriding default settings}
\label{tab:evaluation_params}
\renewcommand{\arraystretch}{1.2}
\resizebox{\linewidth}{!}{%
\begin{tabular}{l l p{4.5cm}}
\hline
\textbf{Parameter} & \textbf{Value} & \textbf{Section} \\
\hline
packet\_size & 128, 512, 1024 & Payload sweep \\
\hline
packet\_size & 512 & All other sections \\
\hline
Clients & 150 & Time-series, Energy \\
\hline
Clients & 100 & Payload sweep \\
\hline
Clients & 80, 150 & PDR heatmap \\
\hline
Clients & 80, 100, 120, 150 & Load sweep \\
\hline
pdr & 0.60 \& 0.80 & 0.80: Load sweep, Time-series, Payload sweep, Energy. Extra graph for PDR\% 0.6 \\
\hline
pdr & 0.50, 0.60, 0.70 & PDR heatmap \\
\hline
burst\_error\_prob & 0.20 (burst) / 0.01 (continuous) & Time-series \\
\hline
burst\_recovery\_prob & 0.10 (burst) / 0.80 (continuous) & Time-series \\
\hline
burst\_loss\_multiplier & 0.40 (burst) / 1.0 (continuous) & Time-series \\
\hline
sim\_time\_s & 300 s & Load sweep \\
\hline
sim\_time\_s & 60 s & Payload sweep, Energy, PDR heatmap \\
\hline
sim\_time\_s & 30--300 s & Time-series \\
\hline
\end{tabular}%
}
\end{table}

\subsection{Numerical Results and Analysis}
This section provides numerical results and analysis on various performance metrics including PDR, goodput, Latency, deadline miss rate along with robustness estimation, energy and sensitivity analysis, and per-attempt ensemble accuracy. 
\subsubsection{Packet Delivery Ratio}
\begin{figure}[t]
\centering
\includegraphics[width=1.0\linewidth]{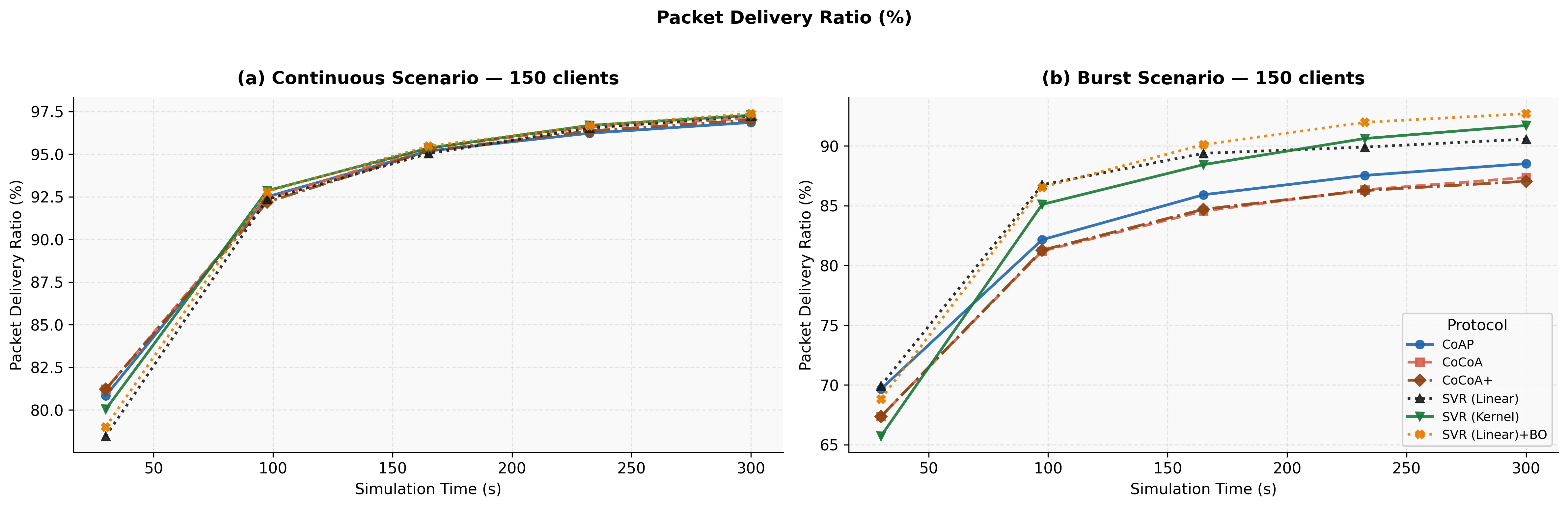}
\caption{ PDR as a function of concurrent clients at PDR = 0.8}
\label{fig:pipe2}
\end{figure}

\begin{figure}[t]
\centering
\includegraphics[width=1.0\linewidth]{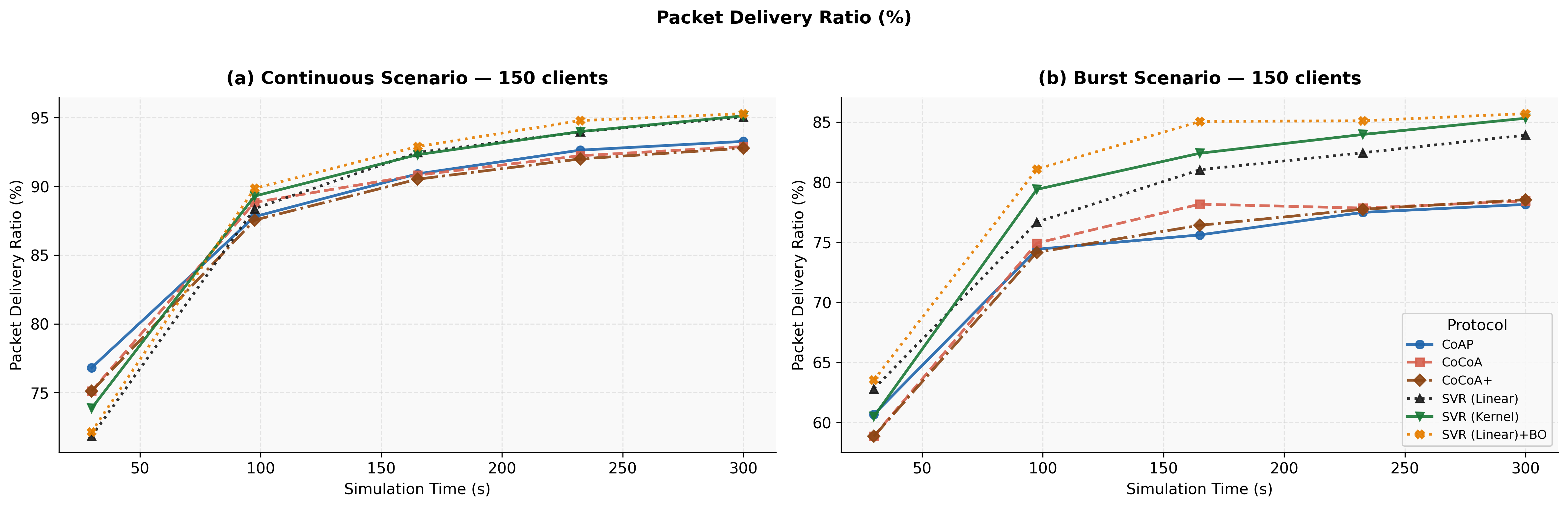}
\caption{ PDR as a function of concurrent clients at PDR = 0.6 }
\label{fig:pipe3}
\end{figure}

Fig. \ref{fig:pipe2} shows the PDR achieved by each protocol under normal conditions (PDR = 0.80). Fig. \ref{fig:pipe3} shows the PDR achieved under stressed conditions (PDR = 0.60) for each protocol. PDR measures the proportion of successfully received packets at the destination compared to those transmitted by the sender \cite{ghebleh2018comparative}. Under a degraded channel with a PDR of 0.60, there is minimal spread among the protocols. Although overall throughput is lower, performance across the protocols remains comparable.

Fig. \ref{fig:pipe3} reports that SVR Linear + BO achieves the highest PDR in the burst scenario, illustrating that the ML model performs strongly in congested channels. Linear + BO is closely followed by Kernel and Linear.

\subsubsection{Goodput}
Goodput is presented in bytes per second as shown in Table~\ref{tab:goodput}, which accounts for the cost of retransmissions. A model that achieves a high PDR through excessive retransmissions rather than efficient RTO management is penalised here, as the associated overhead reduces the effective data rate available to the application layer.
\begin{table}[h]
\centering
\caption{Goodput (B/s) by number of clients across different congestion control and SVR-based approaches.}\label{tab:goodput}
\resizebox{\linewidth}{!}{%
\begin{tabular}{c c c c c c c}
\toprule
\textbf{Clients} & \textbf{CoAP} & \textbf{CoCoA} & \textbf{CoCoA+} & \textbf{SVR (Kernel)} & \textbf{SVR (Linear)} & \textbf{SVR (Linear)+BO} \\
\midrule
80  
& \cellcolor{green!62}8745.0
& \cellcolor{green!63}8755.2
& \cellcolor{green!64}\textbf{8756.9}
& \cellcolor{green!17}8444.6
& \cellcolor{green!5}8357.5
& \cellcolor{green!7}8371.2 \\
100
& \cellcolor{green!70}\textbf{8799.6}
& \cellcolor{green!62}8743.3
& \cellcolor{green!61}8739.8
& \cellcolor{green!29}8519.7
& \cellcolor{green!20}8459.9
& \cellcolor{green!15}8427.5 \\
120
& \cellcolor{green!61}8736.4
& \cellcolor{green!65}\textbf{8768.9}
& \cellcolor{green!65}8763.7
& \cellcolor{green!27}8511.1
& \cellcolor{green!7}8372.9
& \cellcolor{green!12}8408.7 \\
150
& \cellcolor{green!60}8734.7
& \cellcolor{green!65}8765.4
& \cellcolor{green!67}\textbf{8780.8}
& \cellcolor{green!23}8478.7
& \cellcolor{green!5}8326.8
& \cellcolor{green!41}8606.7 \\
\bottomrule
\end{tabular}%
}
\end{table}

\subsubsection{End-to-End Transaction Latency}
Table \ref{tab:latency} reports the mean end-to-end latency, measured from the first transmission attempt to successful ACK receipt. This metric is jointly determined by the number of retransmissions preceding success and the RTT. A protocol that succeeds but applies large RTO values may exhibit disproportionately high latency, as each failed attempt adds a full RTO delay before the next attempt. Due to the lower RTO predictions of the ML model, lower latency would be expected, as the waiting interval preceding each retransmission is shorter. However, as illustrated in Table \ref{tab:latency}, across all load levels the ML models exhibit higher latency and do not report a marginal latency gain under heavier load.
\begin{table}[t]
\centering
\caption{Latency (ms) by Client Count}
\label{tab:latency}
\resizebox{\linewidth}{!}{%
\begin{tabular}{ccccccc}
\toprule
\textbf{Clients} & \textbf{CoAP} & \textbf{CoCoA} & \textbf{CoCoA+} & \textbf{SVR (Kernel)} & \textbf{SVR (Linear)} & \textbf{SVR (Linear)+BO} \\
\midrule
80  & \textbf{3089.9} & 3272.4 & 3271.8 & 3444.0 & 3600.7 & 3527.2 \\
100 & \textbf{4187.4} & 4349.9 & 4389.8 & 4535.8 & 4730.7 & 4627.3 \\
120 & \textbf{5297.8} & 5424.9 & 5503.6 & 5686.3 & 5967.6 & 5752.3 \\
150 & \textbf{6972.2} & 7191.4 & 7193.4 & 7445.4 & 7737.8 & 7418.0 \\
\bottomrule
\end{tabular}%
}
\end{table}

\subsubsection{Deadline Miss Rate}
\begin{figure}[H]
\centering
\includegraphics[width=1.0\linewidth]{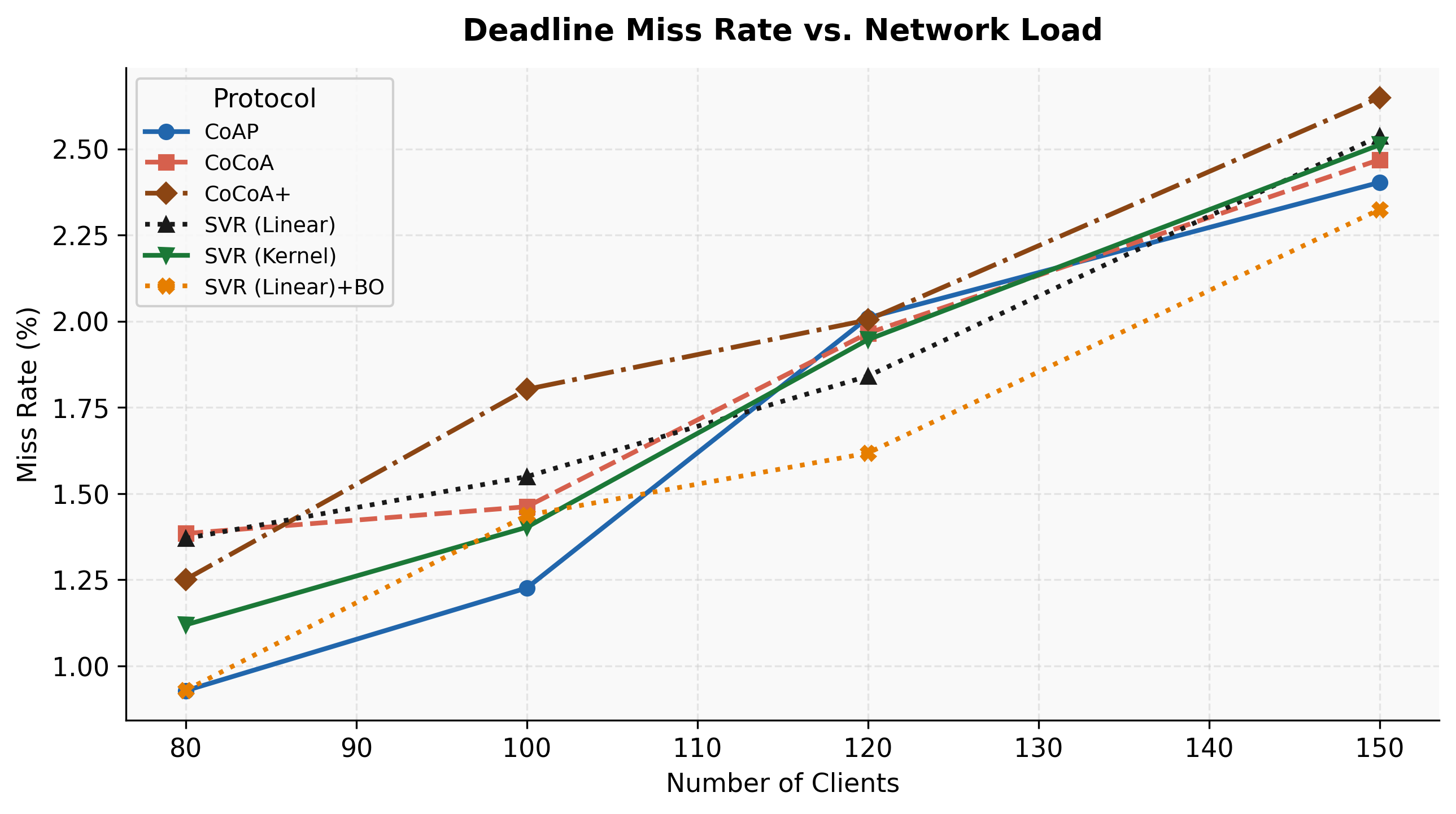}
\caption{Deadline miss rate }
\label{fig:pipe7}
\end{figure}

Fig. \ref{fig:pipe7} reports the delivery miss rate, representing the proportion of packets that are not successfully delivered. This metric is expected to increase with the number of concurrent clients. SVR Linear with BO demonstrates the lowest miss rate across all load levels, indicating its capacity to reduce congestion in the network.

\subsection{Robustness Evaluation}
The following section evaluates the robustness of the protocols under stressed conditions, whereby the PDR is reduced to 0.60, with 150 clients in the network unless otherwise stated.

\subsubsection{Temporal Behavior}
\begin{figure}[t]
\centering
\subfloat[Average retransmissions per transaction]{
  \includegraphics[width=1\columnwidth]{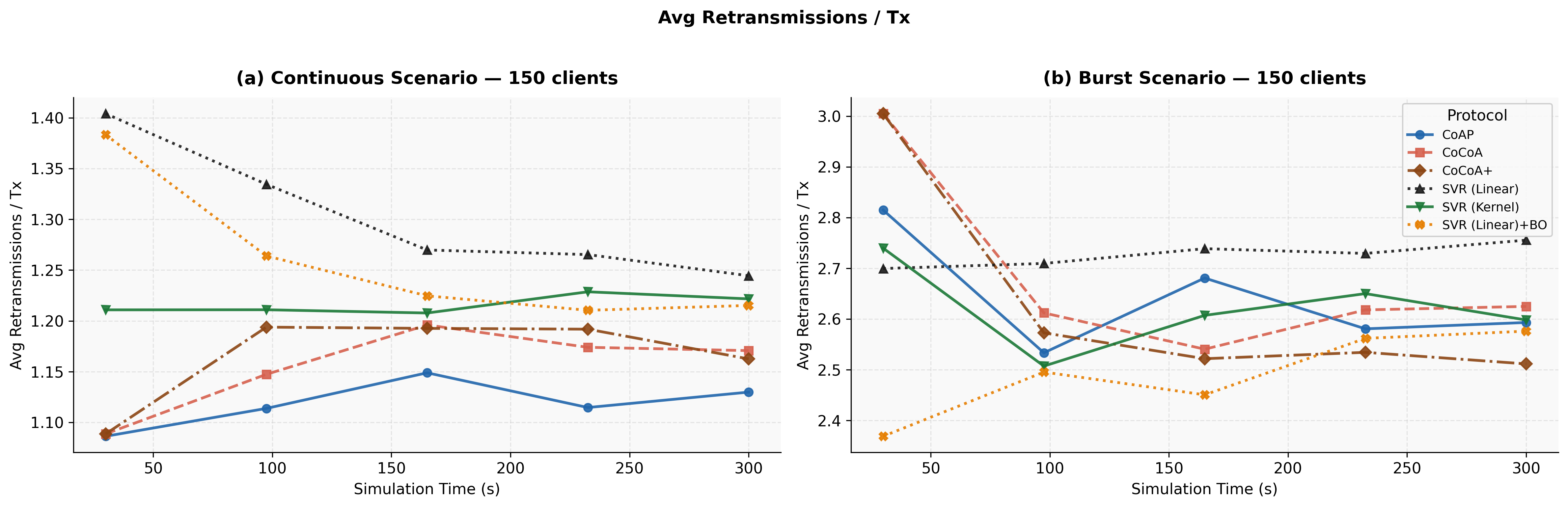}
  \label{fig:pipe8a}
}
\hfill
\subfloat[Average RTO]{
  \includegraphics[width=1\columnwidth]{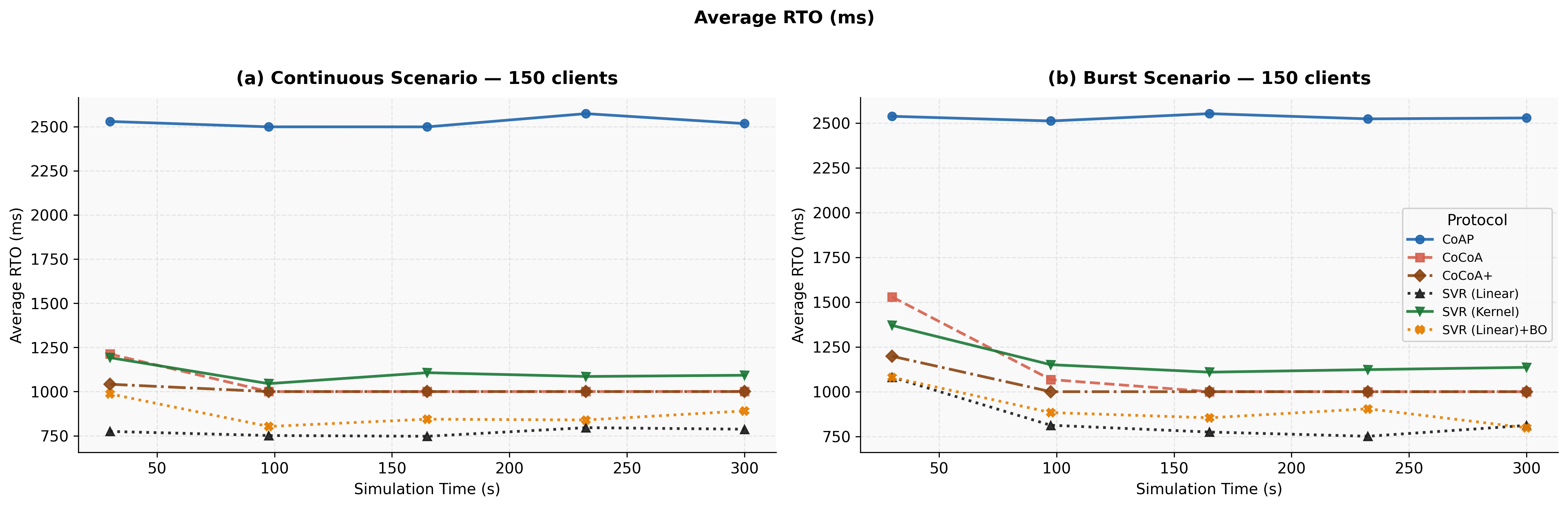}
  \label{fig:pipe9a}
}
\\
\subfloat[Goodput evolution]{
  \includegraphics[width=1\columnwidth]{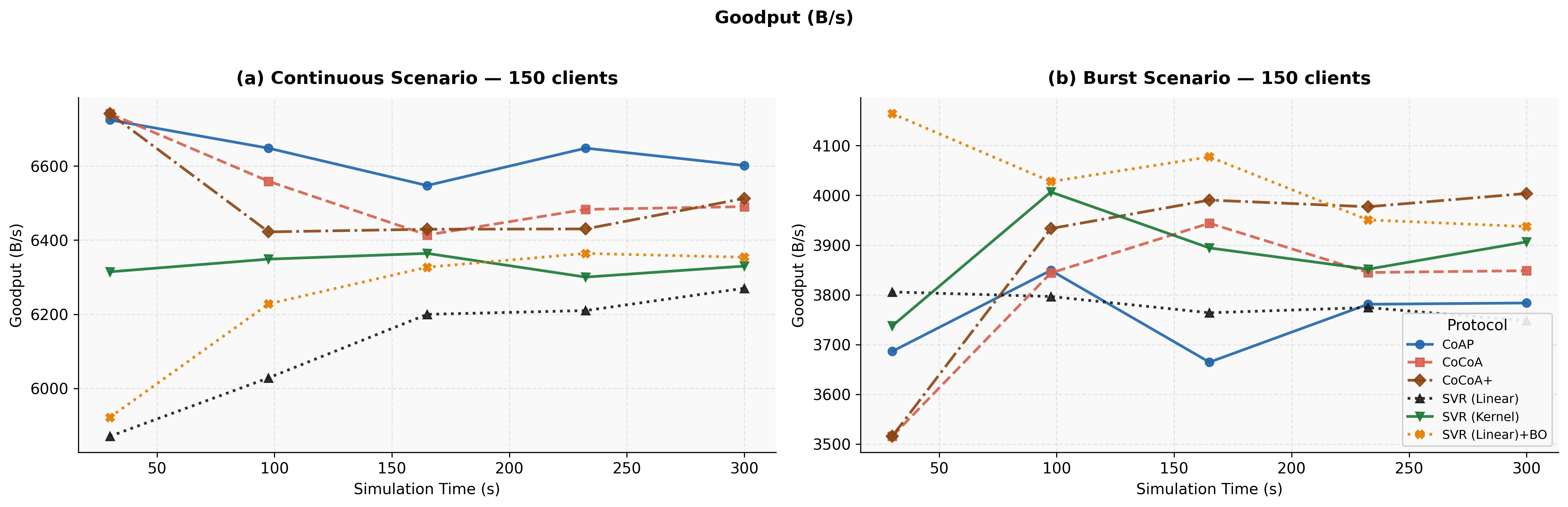}
  \label{fig:pipe10a}
}
\hfill
\subfloat[Cumulative energy consumption]{
  \includegraphics[width=1\columnwidth]{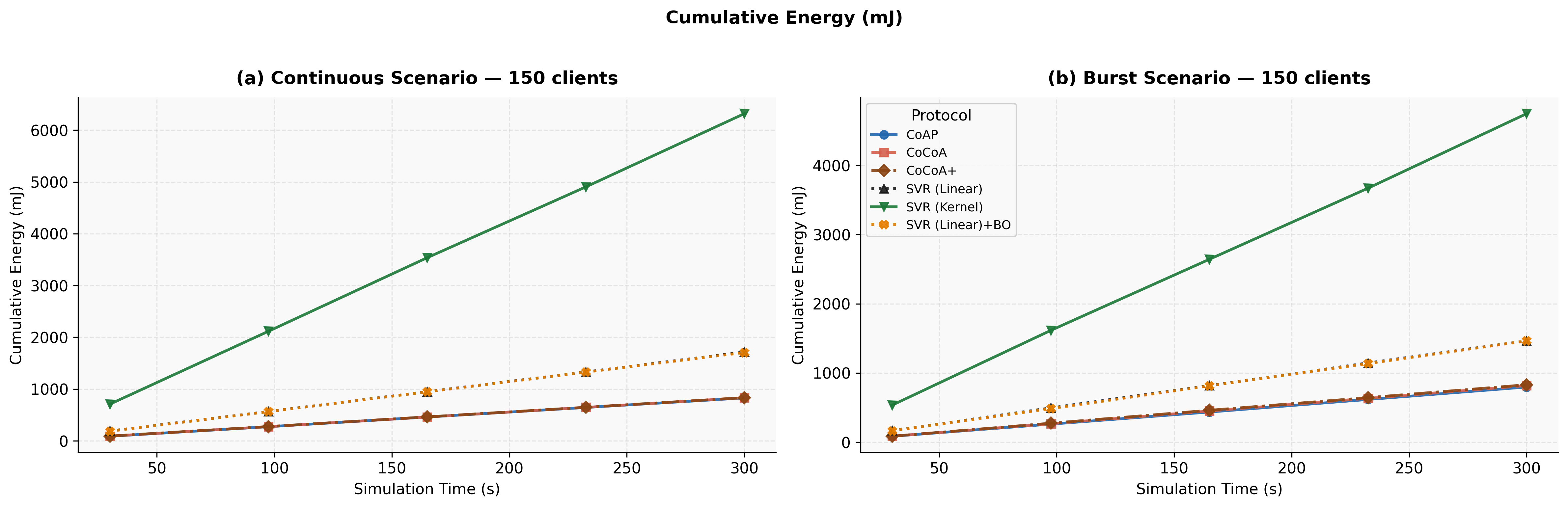}
  \label{fig:pipe11a}
}
\caption{Temporal behavior under link PDR $=0.6$: (a) retransmissions, (b) RTO, (c) goodput, and (d) cumulative energy.}
\label{fig:pipe_time06}
\end{figure}

Fig. \ref{fig:pipe_time06} provides time series for four metrics across two channel conditions. The first represents a continuously degraded channel with \texttt{burst\_error\_prob} = 0.01 and \texttt{burst\_recovery\_prob} = 0.80, corresponding to steady low quality. The second evaluates the model under a bursty channel with \texttt{burst\_error\_prob} = 0.20 and \texttt{burst\_recovery\_prob} = 0.10, characterized by intermittent severe degradation episodes with an average burst duration of 10 simulation steps. Each simulation has a time duration from 30 to 300 seconds. In Fig. \ref{fig:pipe_time06}(a), a lower value is preferable, as it indicates that fewer retransmissions were required per successfully delivered packet.

\subsubsection{Payload Size}
\begin{figure}[!t]
\centering
\includegraphics[width=0.950\linewidth]{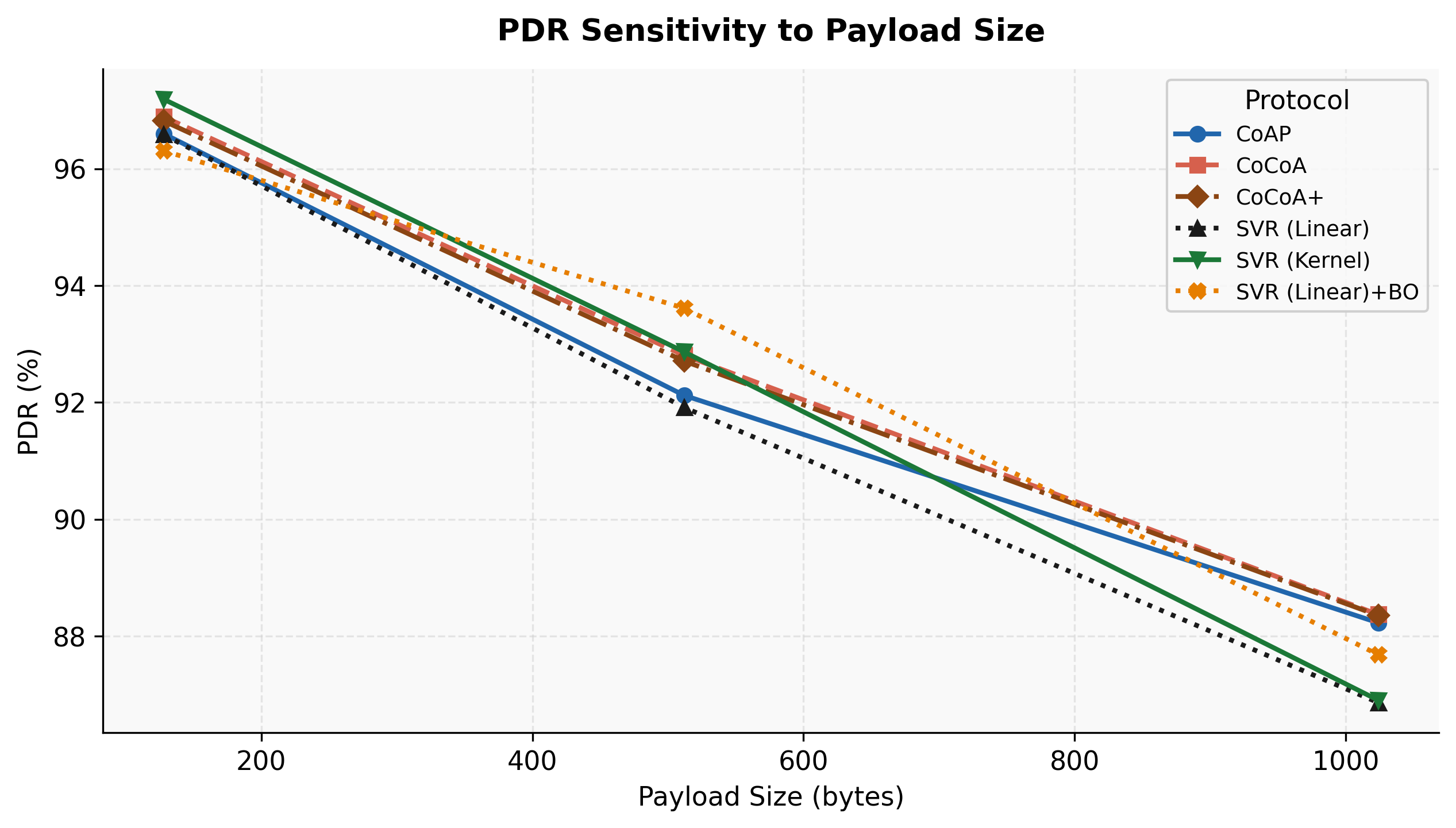}
\caption{PDR as a function of payload size (100-clients, PDR = 0.6)}
\label{fig:pipe12}
\end{figure}

Fig. \ref{fig:pipe12} reports PDR as a function of payload size for 128, 512, and 1,024 bytes. Larger packets occupy the medium for a longer duration, thereby increasing the probability of collisions.

% \subsection{Energy Analysis}
% Here we provide energy consumptions via inference energy breakdown and sensitivity analysis. 

\subsubsection{Radio vs Inference Energy Breakdown}
\begin{figure}[t]
\centering
\includegraphics[width=1.0\linewidth]{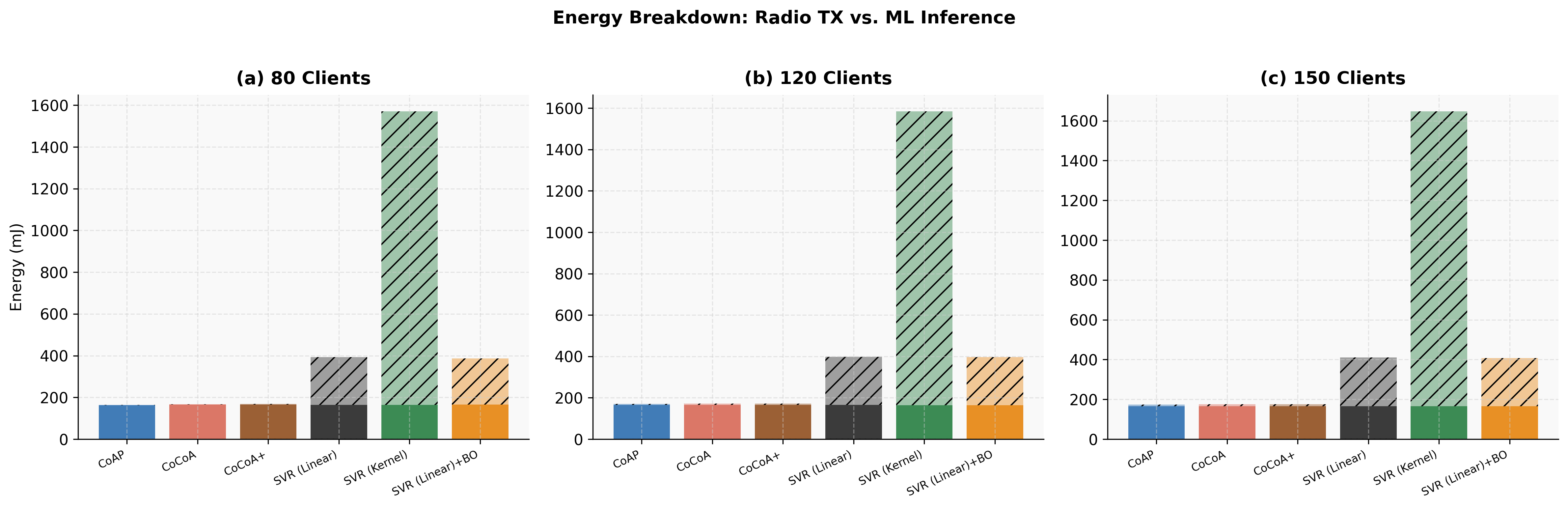}
\caption{Energy breakdown (radio vs ML inference) across three client counts (PDR = 0.8)}
\label{fig:pipe13}
\end{figure}

Fig. \ref{fig:pipe13}(A-C) illustrates the energy breakdown across three network load levels of 80, 120, and 150 clients. Radio transmission energy is represented by the solid fill, while ML inference energy is represented by the hatched fill. The estimated radio energy is 0.1 mJ per packet attempt, based on typical IEEE 802.15.4 transceiver current consumption of 17.4 mA for the CC2420 and packet transmission duration \cite{ti2007cc2420}.

\subsubsection{Sensitivity Analysis}
\begin{figure}[t]
\centering
\includegraphics[width=1.0\linewidth]{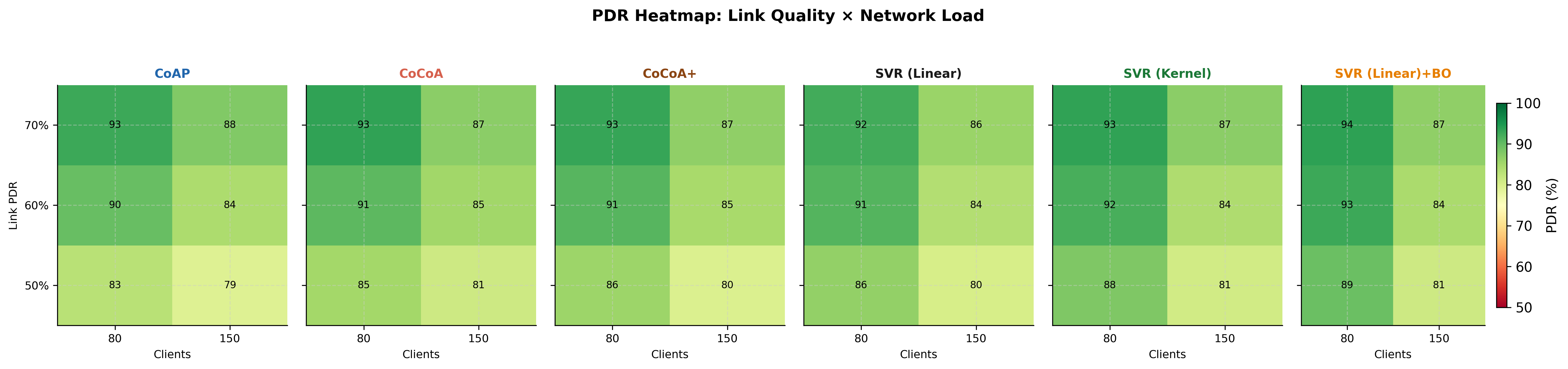}
\caption{PDR heatmap across link quality (rows) and client count (columns) }
\label{fig:pipe14}
\end{figure}

For the vertical axis in Fig.~\ref{fig:pipe14}, the values represent the link-layer PDR settings of 0.50, 0.60, and 0.70, whilst the scale to the right represents the achieved PDR, with each grid cell reporting PDR as a percentage. The heatmap sweeps across 80 and 150 clients. The colour scheme ranges from low PDR, represented in red, to high PDR, represented in green.

\subsection{Per-Attempt Ensemble Accuracy}\label{sec:accuracy}
Table~\ref{tab:linear_svr} summarizes the validation performance of the proposed per-attempt Linear SVR ensemble. We train one sub-model per retransmission attempt index and bound the regression target using the BEB-aligned caps defined by RFC~7252. This decomposition constrains each learner to a narrower, approximately unimodal target interval, which mitigates the limited expressiveness of a linear kernel. The ensemble achieves strong fit for attempts 1--4, with per-attempt $R^2$ values above 0.73. Performance degrades for attempt 5 and beyond ($R^2=0.2856$), which we attribute to the small number of training samples reaching these late retransmission stages. The predicted RTO values remain physically plausible across attempts: the model outputs 610\,ms at attempt~0 and increases to 4{,}048\,ms by attempt~3, staying within the corresponding attempt-specific caps. This behavior indicates adaptation to observed channel conditions rather than a fixed timeout schedule. For reference, the kernel SVR achieves a test $R^2$ of 0.8429 with MAE 80.97\,ms; we include it to quantify the accuracy attainable with a more computationally intensive model.

\begin{table}[!t]
\centering
\caption{Per-attempt Linear SVR performance validation.}
\label{tab:linear_svr}
\resizebox{\columnwidth}{!}{%
\begin{tabular}{r r r c c}
\toprule
\textbf{Attempt} & \textbf{Samples} & \textbf{Cap (ms)} &
\textbf{Test $R^2$} & \textbf{MAE (ms)} \\
\midrule
0   & 859,646 & 2,000  & 0.6384 & 126.4 \\
1   & 379,239 & 4,000  & 0.7766 & 107.8 \\
2   & 187,560 & 8,000  & 0.7712 & 105.8 \\
3   & 98,349  & 16,000 & 0.7636 & 104.5 \\
4   & 51,516  & 32,000 & 0.7297 & 105.4 \\
5+  & 615     & 60,000 & 0.2856 &  81.9 \\
\midrule
\textbf{Overall} & --- & --- & \textbf{0.6303} & \textbf{117.4} \\
\bottomrule
\end{tabular}%
}
\end{table}

\section{Conclusions}\label{sec5}
This paper proposed an on-device lightweight ML framework for CoAP RTO prediction that uses a per-attempt Linear SVR ensemble with a 768-byte footprint and a calibrated RF drop classifier. We evaluated three variations of our approach against CoAP, CoCoA, CoCoA+, and a kernel SVR reference in a validated IEEE 802.15.4 simulator. The Linear SVR improves PDR under the tested load conditions and delivers a 37\% goodput gain under burst traffic. Results also show that the kernel SVR imposes an 84.6\% energy overhead, which indicates that higher nonlinear regression accuracy does not justify its computation cost on constrained devices. The per attempt decomposition follows the BEB attempt scaling and reduces the target range seen by each sub model, which allows a linear predictor to compete with nonlinear alternatives while remaining deployable on resource constrained IoT devices. A key limitation is that training and most evaluation rely on simulator-generated data, so performance may shift under unseen real-world link dynamics and hardware timing effects. In the future, we will extend the drop classifier to cover a wider range of congestion conditions, evaluate on physical hardware at scale, and study XGBoost and TinyML quantization to improve the accuracy and efficiency trade-off.

\bibliographystyle{IEEEtran}
% \clearpage
\bibliography{ref.bib}

@inproceedings{donta2023towards,
  title={Towards Intelligent Data Protocols for the Edge},
  author={Donta, Praveen Kumar and Dustdar, Schahram},
  booktitle={2023 IEEE International Conference on Edge Computing and Communications (EDGE)},
  pages={372--380},
  year={2023},doi={10.1109/EDGE60047.2023.00060},
  organization={IEEE}
}

@article{donta2023icocoa,
  title={{iCoCoA}: intelligent congestion control algorithm for {CoAP} using deep reinforcement learning},
  author={Donta, Praveen Kumar and Srirama, Satish Narayana and Amgoth, Tarachand and Annavarapu, Chandra Sekhara Rao},
  journal={Journal of Ambient Intelligence and Humanized Computing},
  volume={14},
  number={3},
  pages={2951--2966},
  year={2023},doi={10.1007/s12652-023-04534-8},
  publisher={Springer}
}

@article{donta2022survey,
  title={Survey on recent advances in IoT application layer protocols and machine learning scope for research directions},
  author={Donta, Praveen Kumar and Srirama, Satish Narayana and Amgoth, Tarachand and Annavarapu, Chandra Sekhara Rao},
  journal={Digital Communications and Networks},
  volume={8},
  number={5},
  pages={727--744},
  year={2022},
  publisher={Elsevier}
}

@article{Alaa2025towards,
author = {Saleh, Alaa and Morabito, Roberto and Dustdar, Schahram and Tarkoma, Sasu and Pirttikangas, Susanna and Lov\'{e}n, Lauri},
title = {Towards Message Brokers for Generative AI: Survey, Challenges, and Opportunities},
year = {2025},
issue_date = {January 2026},
publisher = {Association for Computing Machinery},
address = {New York, NY, USA},
volume = {58},
number = {1},
issn = {0360-0300},
url = {https://doi.org/10.1145/3742891},
doi = {10.1145/3742891},
journal = {ACM Comput. Surv.},
month = sep,
articleno = {20},
numpages = {37}
}

@article{aveleira2025coap_uad,
  title={CoAP\_UAD: CoAP under attack dataset—A comprehensive dataset for CoAP-based IoT security research},
  author={Aveleira-Mata, Jose and Michelena, {\'A}lvaro and Garc{\'\i}a-Rodr{\'\i}guez, Isa{\'\i}as and Calvo-Rolle, Jos{\'e} Luis and Benavides, Carmen and Jove, Esteban},
  journal={Data in Brief},
  pages={112210},
  year={2025},
  publisher={Elsevier}
}

@article{lopez6721637coap,
  title={CoAP-EAP: A New Standard For Authentication in IoT environments},
  author={Lopez-Gomez, Francisco and Alvarez-Belotto, Ivan and Garcia-Carrillo, Dan and L{\'o}pez-Mill{\'a}n, Gabriel and Mar{\'\i}n-L{\'o}pez, Rafael},
  journal={Available at SSRN 6721637}, year={2026}
}

@article{ghebleh2018comparative,
  author  = {Ghebleh, R.},
  year    = {2018},
  title   = {A comparative classification of information dissemination approaches in vehicular ad hoc networks from distinctive viewpoints: A survey},
  journal = {Computer Networks},
  volume  = {131},
  pages   = {15--37},
  doi     = {10.1016/j.comnet.2017.12.003}
}

@article{betzler2015cocoa,
  author  = {Betzler, A. and Gomez, C. and Demirkol, I. and Paradells, J.},
  title   = {{CoCoA+}: An advanced congestion control mechanism for {CoAP}},
  journal = {Ad Hoc Networks},
  volume  = {33},
  pages   = {126--139},
  year    = {2015},
  doi     = {10.1016/j.adhoc.2015.04.007}
}

@inproceedings{jarvinen2018fasor,
  title={FASOR retransmission timeout and congestion control mechanism for CoAP},
  author={Jarvinen, Ilpo and Raitahila, Iivo and Cao, Zhen and Kojo, Markku},
  booktitle={2018 IEEE Global Communications Conference (GLOBECOM)},
  pages={1--7},
  year={2018},
  organization={IEEE}
}

@article{betzler2016coap,
  author  = {Betzler, A. and Gomez, C. and Demirkol, I. and Paradells, J.},
  title   = {{CoAP} congestion control for the Internet of Things},
  journal = {IEEE Communications Magazine},
  volume  = {54},
  number  = {7},
  pages   = {154--160},
  year    = {2016},
  doi     = {10.1109/MCOM.2016.7509394}
}

@article{akpakwu2020cacc,
  title={{CACC}: Context-aware congestion control approach for lightweight {CoAP/UDP}-based Internet of Things traffic},
  author={Akpakwu, Godfrey A and Hancke, Gerhard P and Abu-Mahfouz, Adnan M},
  journal={Transactions on Emerging Telecommunications Technologies},
  volume={31},
  number={2},
  pages={e3822},
  year={2020},
  publisher={Wiley Online Library}
}

@article{jiang2021machine,
  title={When machine learning meets congestion control: A survey and comparison},
  author={Jiang, Huiling and Li, Qing and Jiang, Yong and Shen, Gengbiao and Sinnott, Richard and Tian, Chen and Xu, Mingwei},
  journal={Computer Networks},
  volume={192},
  pages={108033},
  year={2021},
  publisher={Elsevier}
}

@article{demir2020mlcocoa,
  author  = {Demir, A. K. and Abut, F.},
  title   = {{mlCoCoA}: A machine learning-based congestion control for {CoAP}},
  journal = {Turkish Journal of Electrical Engineering and Computer Sciences},
  volume  = {28},
  number  = {5},
  pages   = {2863--2882},
  year    = {2020},
  doi     = {10.3906/elk-2003-17}
}

@article{li2015internet,
  title={The internet of things: a survey},
  author={Li, Shancang and Xu, Li Da and Zhao, Shanshan},
  journal={Information systems frontiers},
  volume={17},
  number={2},
  pages={243--259},
  year={2015},
  publisher={Springer}
}

@techreport{rfc7228,
  author      = {Bormann, C. and Ersue, M. and Keranen, A.},
  title       = {Terminology for Constrained-Node Networks},
  institution = {IETF},
  number      = {RFC 7228},
  year        = {2014},
  doi         = {10.17487/RFC7228}
}

@manual{ti2007cc2420,
  title        = {{CC2420}: 2.4 {GHz} {IEEE} 802.15.4/{ZigBee}-ready 
                 {RF} transceiver (Rev. C)},
  organization = {Texas Instruments},
  year         = {2007},
  url          = {https://www.ti.com/lit/ds/symlink/cc2420.pdf}
}
\balance
\end{document}